# High-efficiency broadband fiber-to-chip coupler using a 3D nanoprinting microfiber

Dong-Hui Fan,[1,2,3] Xing-Yu Zhang,[1,2] Wei-Jun Zhang,[1,2,3*] Ruo-Yan Ma,[1,2,3] Jia-Min Xiong,[1,2,3] Yu-Ze Wang,[1,2,3] Zhi-Gang Chen,[1,2,3] Zhen Wang,[1,2,3] and Li-Xing You[1,2,3†]

[1]*State Key Lab of Functional Materials for Informatics, Shanghai Institute of Microsystem and Information Technology (SIMIT), Chinese Academy of Sciences (CAS), 865 Changning Rd., Shanghai, 200050, China.*
[2]*CAS Center for Excellence in Superconducting Electronics (CENSE), 865 Changning Rd., Shanghai, 200050, China.*
[3]*Center of Materials Science and Optoelectronics Engineering, University of Chinese Academy of Sciences, Beijing 100049, China.*
[*]zhangweijun@mail.sim.ac.cn; [†]lxyou@mail.sim.ac.cn

**Abstract:** We propose a method for coupling a tapered optical fiber to an inverted tapered SiN waveguide by fabricating a microfiber using 3D nanoprinting lithography. The microfiber consists of three parts: a tapered cladding cap, an S-bend, and a straight part, all composed of high-refractive-index material. Light is adiabatically coupled from the tapered fiber to the printed microfiber through the cladding cap. The light is then transmitted through the S-bend and the straight part with low loss and is finally coupled to the waveguide through the evanescent field. In the simulation, our design can achieve a high coupling efficiency (TE mode) of ~97% at a wavelength of 1542 nm with a wide bandwidth of ~768 nm at the 1-dB cut-off criterion.

1. Introduction

Integrated photonics has attracted much attention because it is a promising high-density integration platform for classical and quantum optical applications in communicating, sensing, and information processing [1]. For these applications, efficient and broadband optical-fiber-to-chip coupling is highly desired for chip packaging and wafer testing. However, realizing such coupling remains a challenge [2], due to the large mismatch in dimensions between the fiber and the waveguide.

A grating coupler (GC) [3,4] is a widely used fiber-to-chip coupling method, but its coupling efficiency is generally low (<50%) [5]. Although its coupling efficiency can be increased to over 80% by using apodization and back reflectors [6,7], these designs are difficult to implement, have limited coupling bandwidth (tens of nm), and are very sensitive to incident angle. An edge coupler [8,9] is superior to GC in terms of bandwidth and coupling efficiency (e.g., >120 nm bandwidth and ~89% at 1550 nm [10]), but for mode-spot matching, its on-chip structure needs to be specially designed, the size of coupler is bulky, and the coupling position is limited.

An alternative coupling method is evanescent coupling with a microfiber [11,12], which does not require phase- and mode-matching, and can achieve high coupling efficiency (>92%), large bandwidth (hundreds of nm), and flexible coupling position [13]. However, because the refractive index of the tapered fiber is less than that of the silica substrate, the on-chip coupling waveguide needs to be suspended during coupling to avoid mode-field leakage to the substrate. This adds great difficulty and packaging inconvenience [14]. In 2020, Khan et al. [15] proposed the use of a high-refractive-index cladding cap on the tapered fiber to overcome the effect of substrate leakage. They fabricated a cladding cap on the tapered fiber by UV exposure and achieved <1.1 dB coupling loss without suspended on-chip structure. However, the cladding cap has to be very carefully designed, and the coupling alignment is not easily scalable.

Recently, three-dimensional (3D) nanoprinting lithography based on two-photon polymerization or two-photon grayscale lithography, has made remarkable achievements in integrated photonics [16–18], due to high *in-situ* lithography precision and a high degree of design freedom. In 2015, Lindenmann et al. demonstrated a photonic wire bonding (PWB) technology with a maximum coupling efficiency of 67% at 1550 nm by using 3D nanoprinting lithography and spot-size conversion [19, 20]. However, the transmission loss due to the mode-field mismatch between the printed PWB and the Si waveguide is still large; for instance, a minimum simulated loss was reported as ~0.75 dB [19] (i.e., a coupling efficiency of ~84%). Further improvement is required.

In this work, we designed and simulated an alternative adiabatic coupler combining tapered optical fiber and 3D nanoprinting technology. The printed coupler (referred to as the microfiber) is composed of a high-refractive-index material that couples the tapered fiber to an on-chip waveguide. The straight section of the microfiber is extended to the surface of the on-chip SiN waveguide, where the light is coupled to the waveguide through the evanescent field. Numerically, we achieve efficient ultra-broadband coupling with a bandwidth of ~768 nm (1-dB cutoff) in a wavelength range of 1432-2200 nm, spanning the main telecom band. The highest coupling efficiency is ~97% at 1542 nm. The on-chip coupling region does not need to be suspended, which simplifies the process. In contrast to the PWB scheme, we use a tapered fiber instead of a 3D nanoprinting structure for spot conversion, which may reduce the partial shadowing effect in the printing process [20]. Furthermore, due to the use of a tapered fiber, the coupling structure that needs to be printed is relatively small, has a fixed configuration, and does not need to be individually designed for each optical fiber, which makes it promising for automated mass manufacturing. The proposed coupler is beneficial for applications in the field that require high efficiency and wide bandwidth, such as optical quantum communication, wavelength division multiplexers, and fiber-to-chip detection.

## 2. Design of coupling structure

Tapered fiber is a mode-field conversion technique [21]. Theoretically, the propagation length of the mode in the tapered fiber needs to be much greater than the beat length $Z_b = 2\pi/(\beta_1 - \beta_2)$, where $\beta_1$ and $\beta_2$ are, respectively, the propagation constants for the fundamental and higher-order modes [22]. There are many ways to realize a tapered fiber, such as wet etching and fusion tapering [23,24]. Wet etching is more convenient for controlling the length of the tapered region (~2 mm) and the shape of the terminated facet.

Fig. 1 shows the schematic design of the microfiber. It consists of three parts: a tapered cladding cap (referred as to the "cap"), an S-bend, and a straight part. The printing material IP-DIP has a high refractive index of $n_{\text{IP-DIP}} \sim 1.52$ at 1550 nm [25], higher than that of the silica substrate, $n_{\text{SiO2}} \sim 1.45$ at 1550 nm. The cap wrapped on the surface of the tapered fiber is used to adiabatically channel the light field from the fiber into the microfiber. The small size and the tapered shape of the fiber tip may help reduce the partial shadowing effect of the printing that appears in vertical device facets. The microfiber extends to the top of the inverted tapered SiN waveguide structure for evanescent coupling. To enhance mechanical strength and stability, the coupling region needs to be covered with a low-refractive-index matching adhesive—for example, a UV-curable glue (MY-133EA, My polymers Ltd., $n_{\text{glud}} \sim 1.33$ at 1550 nm), which has good stability at a cryogenic temperature [26,27]. Thus, it is possible to use this method for cryogenic applications, such as the efficient coupling of the fiber to a waveguide-integrated superconducting single-photon detector [28].

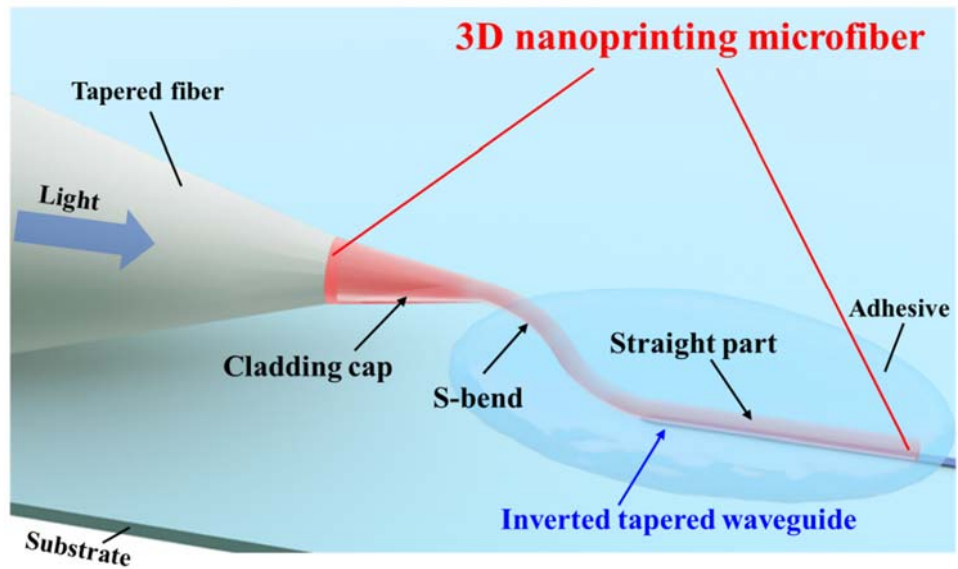

**Figure 1.** Schematic of evanescent coupling for a tapered fiber to an inverted tapered on-chip SiN waveguide. The three key components of the 3D nanoprinting microfiber are marked: a cladding cap, an S-bend, and a straight part.

## 3. Optimization of the parameters of the coupling structures

## 3.1 Design of the adiabatic coupling cap

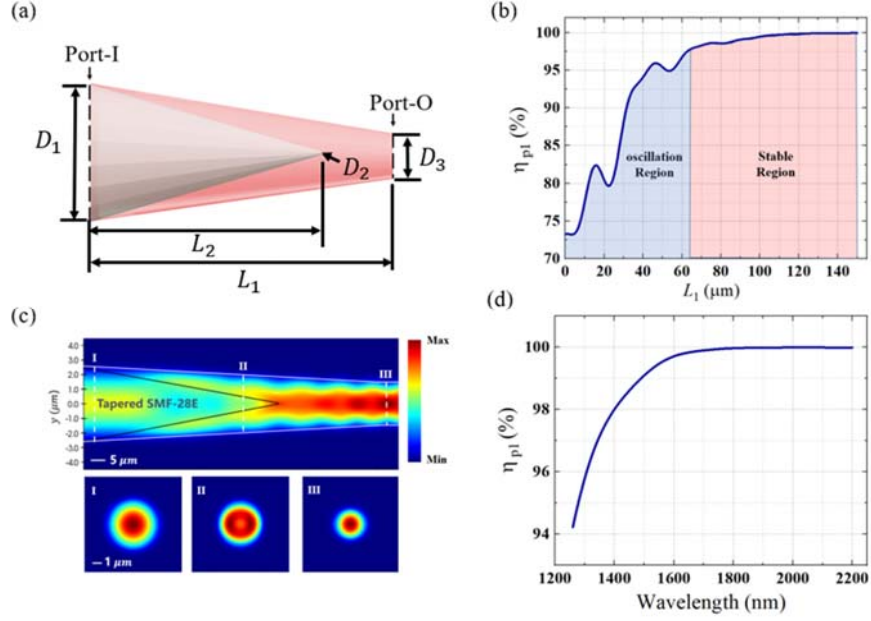

**Figure 2.** (a) Cross-sectional view of the coupling region of the tapered fiber and the cap (red). The design parameters and the port setup in the simulation model are indicated. (b) Coupling efficiency as a function of the coupling length at 1550 nm. (c) Cross-section of electric field distribution in coupling mode when $L_1$ = 100 μm. Three different locations in the coupling region are marked: the input part (Region I, TE fundamental mode), the middle part (Region II, adiabatic coupling), and the output end (Region III, the output mode of the cap). The distribution of the electric field is concentrated in the cap. (d) Coupling efficiency $\eta_{p1}$ as a function of wavelength.

We use the commercial software Lumerical's EME (Eigenmode Expansion) solver [29] to numerically optimize the structure of the adiabatic coupling cap. EME provides a strict solution of Maxwell's equations and is suitable for dealing with lengthy 3D structures. We adopted a tapered SMF-28e fiber in the simulation, which is often used at telecom wavelengths. Fig. 2(a) shows the cross-sectional parameters of the cap. Here $D_1$ is the fiber diameter when the cap begins to wrap the tapered fiber; $D_2$ is the diameter of the tip of the tapered fiber, which is set to zero for simplification; $D_3$ is the diameter of the tip of the cap, after which the diameter remains constant; $L_1$ is the length of the entire cap; and $L_2$ is to the length of the tapered fiber wrapped by the cap.

To achieve a compact coupling structure and reduce the dependence on the printing precision, we set $D_1$ = 5 μm and $D_3$ = 3 μm. The refractive index of the optical fiber used is $n_{fiber}$ = 1.45. In the simulation, we set the incident port (Port-I) at $D_1$ and the output port (Port-O) at $D_3$. The input mode is the fundamental TE mode at a wavelength of 1550 nm.

For the adiabatic cap, the greatest concern is whether the length $L_1$ can achieve adiabatic coupling. As shown in the blue area of Fig. 2(b), the coupling efficiency at first increases slowly and oscillates as $L_1$ increases. This oscillation is caused by the inter-modal interference of the higher-order mode, which is converted from the fundamental mode in the non-adiabatic coupling of the coupled structure. When $L_1 \geq 65$ μm, it is observed that the oscillation of the coupling efficiency decreases significantly, and finally stabilizes at ~ 99% when $L_1 = 100$ μm. Thus, in the subsequent simulations, we fixed $L_1$ at 100 μm, which corresponds to $L_2 = 66$ μm.

We further determined the mode transmission characteristics of the coupling structure. In practical applications, the fundamental mode is preferred. Fig. 2(c) shows a cross-sectional view of the cap along the direction of beam propagation. At the beginning of the coupling structure (Region I), the light field does not directly enter the cap, but gradually couples into the cap as the diameter of the tapered fiber decreases, and the dielectric in the cladding (Region II) acts as a new medium to propagate the mode field. In Region III, the output mode is still the TE fundamental mode, which indicates that the tapered coupling structure can indeed transmit the light field adiabatically and does not couple to higher-order modes.

Fig. 2(d) shows the coupling efficiency of the cap as a function of wavelength. It increases gradually as wavelength increases. This is because the greater the wavelength, the larger the evanescent field formed for the same fiber diameter, and the easier it is for the light field in the tapered fiber to be coupled into the highly refractive index cap, which is equivalent to increasing the cap coupling length. Here we define the wavelength-dependent coupling efficiency of this component as $\eta_{p1}$.

3.2 Design of the S-bend

The tapered fiber is placed horizontally on the substrate as shown in in Figure 1. A height difference occurs between the cap tip and the on-chip waveguide, which is the sum of the fiber radius (~62.5 μm) and the SiN waveguide thickness (~330 nm, negligible here). The two components are connected through the S-bend bridge, which is composed of two arcs, reversely spliced together. Fig. 3(a) shows the design parameters of the S-bend, where $r$ is the bend radius, $\theta$ is the angle of each arc, and the diameter of the fiber is the same as the tip of the cap ($D_3$). To simplify the simulation, 2D-FDTD was used for Fig. 3(b) to determine the parameter range, and 3D-FDTD was used for Fig. 3(c) for the single parameter wavelength simulation.

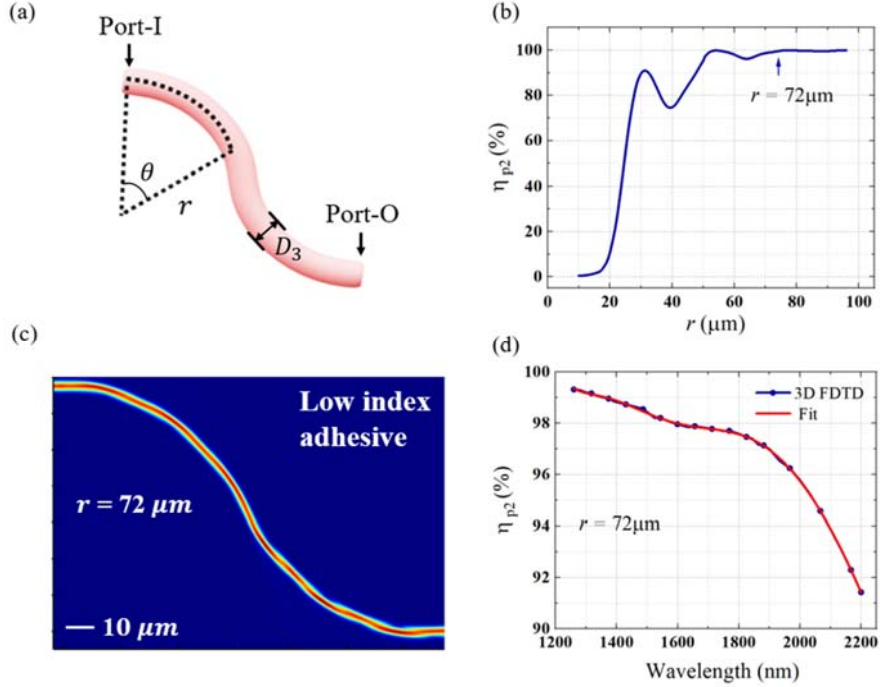

**Figure 3.** (a) Side view of the S-bend, with the design parameters and the port setup used in the simulation. (b) Coupling efficiency $\eta_{p2}$ as a function of bend radius $r$, calculated at 1550 nm. (c) Cross-sectional view of electric field distribution in the S-bend. (d) Wavelength dependence of coupling efficiency $\eta_{p2}$. The red line is a polynomial fit for the simulated data.

Fig. 3(b) shows the simulated transmission of the TE fundamental mode in the S-bend as a function of $r$ at 1550 nm. When $r > 72$ μm, the transmission efficiency can reach more than 99%. Similarly, we scanned the transmission as a function of wavelengths under this optimal bending radius and defined the coupling efficiency as $\eta_{p2}$.

Fig. 3(c) shows the cross-section of the simulated TE fundamental mode in the S-bend with $r = 72$ μm, and $\theta = 0.986$ rad. The light field is completely confined within the S-bend. Fig. 3(d) shows $\eta_{p2}$ as a function of wavelength. In the depicted wavelength range, 1260-2200 nm, $\eta_{p2}$ decreases steadily as wavelength increases. $\eta_{p2}$ is 99.3% at a wavelength of 1260 nm and diminishes to 91.4% at a wavelength of 2200 nm. Because of the decrease of the mode refractive index caused by the increase of the wavelength, the light field leaks to the outside during transmission through the S-bend.

3.3 Design of the straight part

The straight part of the microfiber extends horizontally over the SiN waveguide ($n_{SiN} = 1.99$ at 1550 nm [30]). We define the coupling efficiency of this part as $\eta_{p3}$. When two waveguides are

6 / 14

close together, strong-coupling theory [31] can be used for qualitative analysis. But for specific parameter design, we again used EME for simulation.

Fig. 4(a) shows the design parameters of the straight part, where $w_1$ is the starting width of the SiN inverted tapered waveguide structure, $w_2$ is the final width of the waveguide, and $L_3$ is the length of the coupling between the microfiber and the SiN waveguide. For low-loss coupling, the effective refractive index of the waveguide needs to adiabatically sweep the fundamental mode effective refractive index of the microfiber. Fig. 4(b) shows the cross-sectional view of the coupling region of the straight part and the waveguide.

In the evanescent coupling of microfiber and waveguide, $L_3$ and $w_1$ are the most important parameters. When both $w_1$ and $w_2$ are determined, $L_3$ will affect the taper of the cone structure. The size of the taper determines whether adiabatic coupling can be performed, and $w_1$ determines the initial value of the effective refractive index of the waveguide. Therefore, at a wavelength of 1550 nm, we simulated $\eta_{p3}$ as functions of both $w_1$ and $L_3$. For high-density integration, the coupler should be compact, and the minimum line width should be easy to fabricate. Following this rule, we found that the optimal $w_1$ is 400 nm, and the minimum $L_3$ is 155 μm, as shown in Fig. 4(c). In addition, it is also observed that as $L_3$ increases, the overall coupling efficiency increases.

Fig. 4(d) shows the simulated bandwidth characteristics of the coupled structure. Its coupling efficiency is asymmetric as a function of wavelength, which is different from the symmetric function of the grating couplers at a certain central wavelength [3]. This is caused by the change in the refractive index of the TE fundamental mode due to the wavelength change. The shorter the wavelength, the higher the effective refractive index of the fundamental mode, which is equivalent to a continuous increase of $w_1$. When the propagation constant difference $\Delta\beta$ between the fiber fundamental mode and the waveguide fundamental mode at the beginning of the coupling region exceeds a certain range, coupling efficiency will drop. The decrease in coupling efficiency at long wavelengths is due to the decrease in the effective refractive index of the TE fundamental mode in the SiN waveguide, resulting in mode leakage to the substrate. In addition, the coupling efficiency oscillates with wavelength owing to the energy exchange between the two waveguides occurring through interference. The inset of Fig. 4(d) shows the cross-section of the electric field distribution in the coupling mode, indicating that light-field energy is gradually coupled into the waveguide from the microfiber.

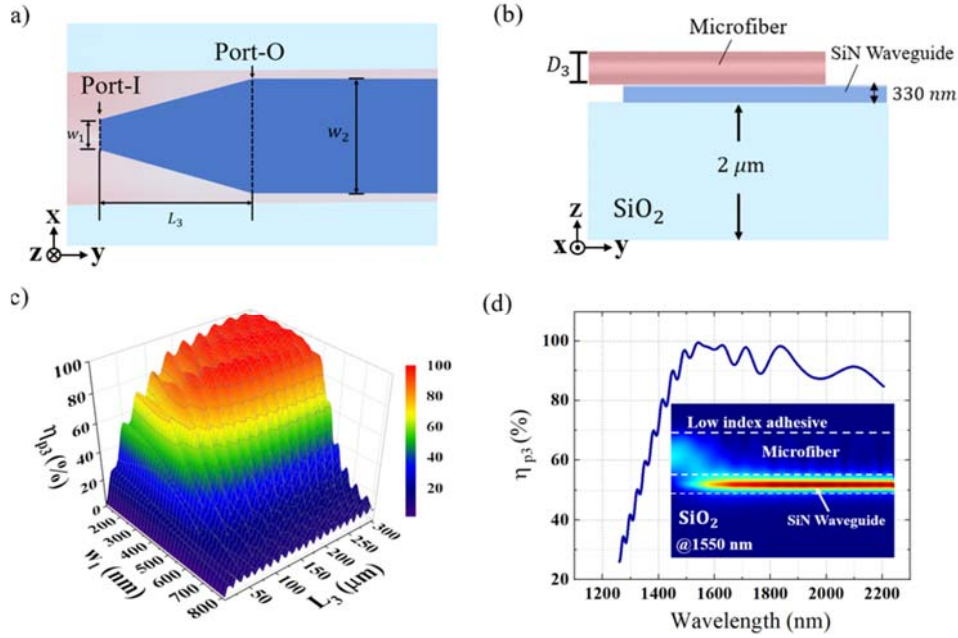

**Figure 4.** (a) Bottom view of the inverted tapered SiN waveguide (dark blue) and the straight part (red). The design parameters and the port setup in the simulation are marked. (b) Cross-sectional view of the coupling region of the straight part and the waveguide. (c) Effect of $L_3$ and $w_1$ on coupling efficiency. (d) Wavelength dependence of the coupling efficiency $\eta_{p3}$. Inset: cross section of electric field distribution in coupling mode, when $L_3$ = 155 μm and $w_1$ = 400 nm.

3.4 Global coupling efficiency calculation

According to the simulations described above, we obtained the optimal design parameters of each part, as well as their coupling efficiencies, as functions of wavelength. We then performed a calculation on the overall efficiency. We call this overall efficiency $\eta_{total}$; its expression is as follows: $\eta_{total} = \eta_{p1} \times \eta_{p2} \times \eta_{p3}$.

Fig. 5 shows $\eta_{total}$ as a function of wavelength. $\eta_{total}$ inherits the curve characteristics of the efficiency of each component to a certain extent. It is observed that $\eta_{total}$ can reach more than 1 dB in the wavelength range 1432-2200 nm, and that the bandwidth is ~768 nm. $\eta_{total}$ reaches a maximum of ~97.1% at 1542 nm and is ~95% at 1550 nm. The non-unity coupling efficiency in this simulation is limited by energy exchange loss between the microfiber and the SiN waveguide.

Note that the above results do not take into account the effects of the transmission mode and the loss in the fiber used. For example, in an SMF-28e fiber, the single-mode operating wavelength range is 1260-1700 nm, and the transmission loss is negligible for a several-meter-long fiber. However, at wavelengths below 1260 nm and above 1700 nm, actual loss may increase and may need to be carefully evaluated, owing to changes in the mode or increase of the intrinsic loss in the fiber.

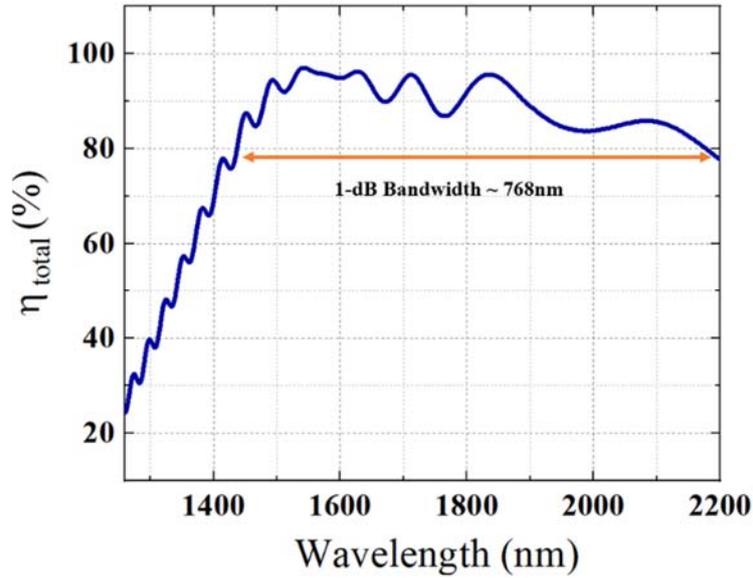

**Figure 5.** Wavelength dependence of the total coupling efficiency $\eta_{total}$.

3.4 Analysis of fabricating tolerance

Here we analyze the tolerance of the proposed coupler to the fabrication errors. First is the fabrication error of the taper fiber. In the actual fabrication the tip diameter ($D_2$) of the taper fiber is less than 600 nm for conventional wet etching [23]. Fig. 6(a) shows that the coupling efficiency is greater than 99% for $D_2 < 600$ nm, and that the efficiency loss caused by the $D_2$ deviations can basically be ignored. Second is the fabrication error on the tip of the inverted tapered waveguide. As shown in Fig. 6(b), with a deviation of ±65 nm, $\eta_{p3}$ is still greater than 1 dB cutoff point. This deviation tolerance is large enough for the typical fabrication precision (±10 nm) of electron beam lithography (EBL) [32].

Generally, the fabrication precision of typical 3D nanoprinting equipment (e.g., Quantum X align, Nanoscribe, Ltd.) is less than ±20 nm [33], and the alignment error is ∼ ±100 nm. Figs. 6(d) and (e) show the variations of the efficiency for different values of $D_3$ and $D_2$. The errors caused by the limited precision of the printing lithography result in insignificant changes in coupling efficiency.

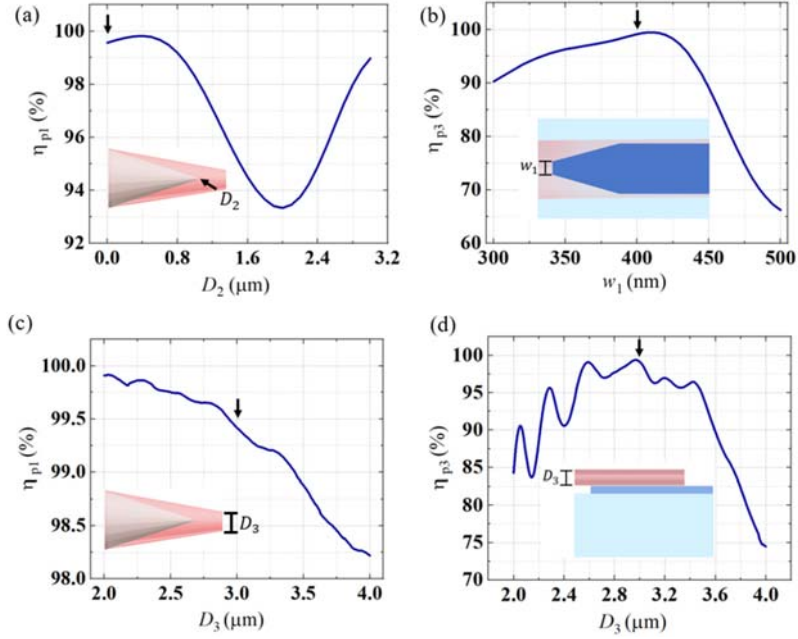

**Figure 6.** Design parameter sweeping for different coupling efficiencies. (a) $\eta_{p1}$ vs. $D_2$. (b) $\eta_{p3}$ vs. $w_1$. (c) $\eta_{p1}$ vs. $D_3$. (d) $\eta_{p3}$ vs. $D_3$. The down arrows mark the parameters used in the simulation in Fig. 5.

Alignment error can occur during the printing, for example between the tapered fiber and the cap, and between the straight part of the microfiber and the waveguide. Figs. 7(a) and (b) show the effect of misalignment between the straight part of the microfiber and the waveguide in $x$- and $z$-directions, respectively. Without including the resonance dips in the $\eta_{p3}$ efficiency curve, the 1-dB offset tolerance in the $x$-direction is ~ 1.56 μm. However, the $z$-direction misalignment is most noticeable in the presented method. Fig. 7(b) shows that the 1-dB offset tolerance is ~200 nm. However, considering that the alignment precision is ~ ± 100 nm, the $z$-direction offset is controllable.

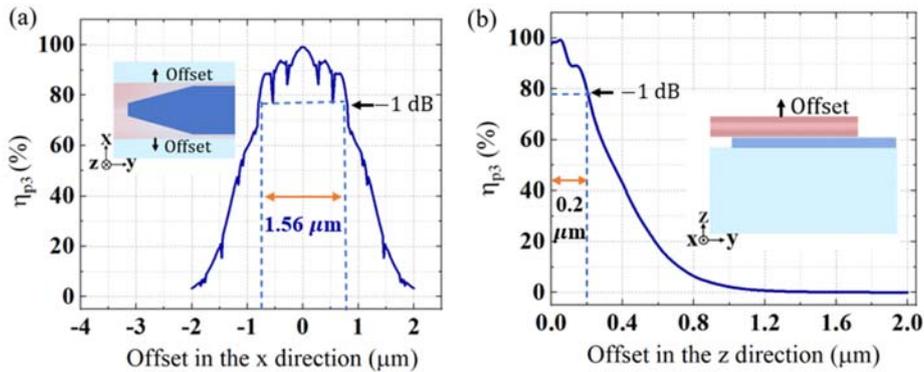

**Figure 7.** Alignment offset tolerance for different cases. (a) $\eta_{p3}$ vs. offset in the $x$-direction. (b) $\eta_{p3}$ vs. offset in the $z$-direction.

4. **Discussion**

Table 1 compares features of our proposal with those of other types of couplers. Our proposal demonstrates a remarkably wide bandwidth, combined with high coupling efficiency and flexible coupling position. Furthermore, the evanescent coupling in our proposal may be robust to the horizontal shrinkage of the resist material during sample development, which has proven to be a serious obstacle in PWB technology [20]. In our proposal, the horizontal shrinkage (i.e., in the $y$-direction in Fig. 4) for the evanescent coupling can be compensated by designing a longer coupling length.

There are some possible ways to improve the present design. For example, the intrinsic loss problem of silica fiber when the wavelength is greater than 1700 nm can be addressed by replacing the silica fiber with an alternate fiber (e.g., a SM-1950 Ge-doped fiber, operating wavelength range of 1850-2200 nm, Thorlabs, Inc.). The design of the printed coupler can then be altered to match the new fiber flexibility in the coupling position.

**Table 1.** Comparison with different couplers in telecom C-band.

| Type of couplers | coupling efficiency (%) | | 1-dB bandwidth (nm) | | 1-dB alignment tolerance (μm) | coupling position |
|---|---|---|---|---|---|---|
| | simulated | measured | simulated | measured | | |
| Conventional GC [34] | ~37 | ~31 | ~47 | ~45 | ~5 | arbitrary |
| Apodized GC (with mirror) [7] | ~90.5 | ~86.7 | ~76 | ~38 | N/A | arbitrary |
| Edge coupler (suspended) [10] | ~93.5 | ~89 | >120 (<-0.5 dB) | >100 | N/A | edge |
| PWB coupler [19] | ~84.1 | ~67 | >100 | N/A | ~2 | edge |
| 3D polymer coupler (micro-reflector) [35] | ~90 | ~90 | >500 (<-0.7 dB) | >300 | ~4.5 | edge |
| This work (simulation) | ~96.5 | N/A | ~768 | N/A | ~1.56 | arbitrary |

Also, a high fabrication precision of EBL can help relax the restrictions on the design parameters of the inverted tapered SiN waveguide. For example, if $w_1$ is reduced from 400 to 300 nm while $L_3$ is maintained unchanged, the simulated 1-dB cutoff bandwidth increases from ~768 to ~884 nm. Although at $w_1 = 300$ nm, the coupling efficiency at 1550 nm decreases, increasing the length of $L_3$ can solve this problem, as illustrated in Fig. 4(c).

For the S-bend, if the height difference between the tip of the tapered fiber and the waveguide can be eliminated, then the microfiber coupler can be made straight—that is, there is no need to design an S-bend. For example, Fig. 8 shows an array of on-chip couplers composed of straight microfibers. This is realized by fabricating a V-groove on the chip to reduce the height difference on the chip. With the V-groove, the orientations between the tapered fibers and the on-chip waveguides are fixed, which provides convenience and coupling precision for subsequent 3D nanoprinting. Because the presented coupling structure provides simple processing and has a compact minimum feature size, it is promising for realizing a large-scale coupling array for practical use. Moreover, the transmission direction of the design has good reversibility—that is, on-chip waveguide coupling to the optical fiber is also feasible.

Finally, compared with a Si-on-insulator platform [3], a SiN waveguide has natural disadvantages in grating coupling because of its low refractive index. Our scheme provides another option for SiN waveguide coupling. This coupling method can be applied to many other materials with a higher refractive index than that of optical fibers. The simple processing structure and large tolerance of this solution is expected to bring great convenience to the design.

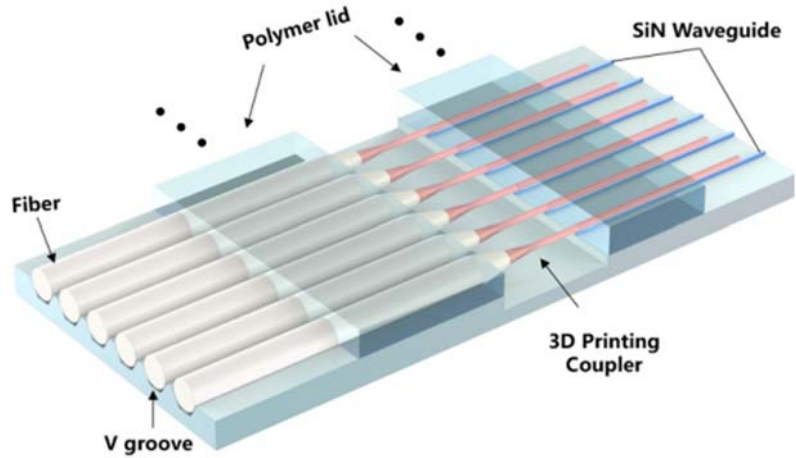

**Figure 8.** Conceptual drawing of arrayed 3D nanoprinting couplers based on tapered fibers.

## 5. Conclusions

In this work, we numerically simulated a fiber-to-chip waveguide coupler based on a tapered fiber. In the simulation, we designed and optimized the geometric parameters and transmitting efficiency of a 3D nanoprinting high-refractive index microfiber, including a tapered cladding cap, an S-bend, and a straight part. It is found that, without the need for suspending the on-chip waveguide structure, a coupling efficiency of ~97% can be achieved at a wavelength of 1542 nm and a 1 dB cut-off bandwidth of ~768 nm in the wavelength range 1432-2200 nm. This study offers an alternative method to reduce the design and fabrication difficulties of the current PWB technology. The proposed coupler has the advantages of broad bandwidth, high coupling efficiency, and simple processing, and is very attractive in the fields of on-chip quantum communication and waveguide-integrated superconducting single-photon detectors. Our results can also provide useful information for studies of evanescent coupling structures.

**Finding**

This work is supported by the National Natural Science Foundation of China (NSFC, Grant No. 61971409), the National Key R&D Program of China (Grants No. 2017YFA0304000), and the Science and Technology Commission of Shanghai Municipality (Grant No. 18511110202, and No. 2019SHZDZX01). W.-J. Zhang is supported by the Youth Innovation Promotion Association